\documentclass[a4paper,10pt]{article}

\usepackage[utf8]{inputenc}
\usepackage{textcomp}
\usepackage{hyperref}
\hypersetup{
    colorlinks=true,
    linkcolor=blue,
    filecolor=magenta,      
    urlcolor=cyan,
}
\usepackage{xcolor}
\usepackage{listings}
\usepackage[cache=true,frozencache,cachedir=.]{minted}
\usepackage{upquote}
\usepackage{graphicx}
\usepackage{amsmath}
\usepackage{tikz}
\usetikzlibrary{calc,patterns,decorations.pathmorphing,decorations.markings}

\title{A practical guide to using the AngCor package}
\author{Philip Adsley, iThemba LABS/University of the Witwatersrand\footnote{philip.adsley@wits.ac.za}\\
Kevin CW Li, iThemba LABS/University of Stellenbosch\\
Harshna Jivan, University of the Witswatersrand\\
Luke Morris, University of York\\
Luna Pellegri, iThemba LABS/University of the Witwatersrand}
\date{\today}

\begin{document}
\lstset{language=bash}
\maketitle

\section{Introduction}

This report contains brief, practically oriented instructions for how to use the various codes contained within the {\it AngCor} software package \cite{AngCorGithubLink}. Most of these codes (everything except the averaging code) have been written by other people. While the contributors to this report may be able to assist users with various parts of the codes, some queries may be beyond our knowledge. Please leave a comment on the \href{https://github.com/padsley/AngCorPackage}{Github repository} if you require assistance.

This guide is a {\it living document} which has been uploaded onto the arXiv to ensure that there is an accessible version of the report which may be used as a static reference for users of the {\it AngCor} package. This report may contain errors. If you find an error please let us know so the guide may be corrected. In addition, if you have any comments, suggestions or corrections, please get in touch so they can be added to the guide.

A final note of warning - we have used this code for various scenarios for experiments performed using the K600 magnetic spectrometer and its ancillary detectors at iThemba LABS, South Africa . If you are trying to use this code with some other experimental setup you may have to make some adjustments to the inputs or the averaging code. If you are doing this and you need help, advice or emotional support, please get in touch and we will do what we can to help.

\section{The Physical Problem}

Angular correlations in nuclear reactions are an important tool in helping to assign spins and parities to states, and in properly computing branching ratios for excited states in nuclei by accounting for the missing solid angle in each case. The angular correlations typically depend on the reaction used to populate the state of interest, the spin and parity, and the mode of decay.

To calculate the form of the decay, typically one needs to have the initial spin substate ($m$-state) distribution. This depends on the reaction and, in the present case, are extracted from the {\it CHUCK3} code. Once the initial $m$-state distribution is known they can be used along with the appropriate decay amplitudes for the reaction of interest to calculate the angular correlations - this is done with the {\it AngCor} code.

Finally, in many cases that are investigated with the K600, the finite acceptance of the spectrometer means that it is necessary to smear out the angular correlations with the appropriate range of recoil angles taking into account the angular acceptance of the spectrometer and the reaction kinematics. This smearing of the angular correlation is done with the averaging portion of the code. This code averages the {\it AngCor} outputs over the reaction region of interest, weights correctly according to the differential cross section of the initial reaction and produces an averaged (observed) angular correlation function. This averaging code has been written so that the output is in the form of a ROOT TTree, meaning that dynamic gates may be placed on the spectrometer aperture.

Recent experimental studies which have used angular correlations of the type described within this work are:

\begin{enumerate}
 \item studies of the PDR using the K600 and BaGeL (Ball of Germanium and LaBr detectors), an array of HPGe and lanthanum bromide detectors;
 \item studies of nuclear clustering in light nuclei such as $^{16}$O \cite{PhysRevC.95.031302} and $^{24}$Mg using the K600 \cite{NEVELING201129} and CAKE (the Coincidence Array for K600 Experiments), an array of silicon detectors \cite{Adsley_2017};
 \item studies of $\gamma$-ray transitions within the candidate superdeformed band in $^{28}$Si using the Grand Raiden magnetic spectrometer and \href{http://www.rcnp.osaka-u.ac.jp/Divisions/np1-a/CAGRA/index.html}{CAGRA} (Clover Array Gamma-ray spectrometer at RCNP/RIBF for Advanced research).
\end{enumerate}
  
If you use this software package to calculate angular correlations, please let us know so we can add a reference to the paper in the above list. In addition, if you are willing, please send us the input files that you have used for your calculations - having reference calculations for different examples is extremely helpful for future users.

\section{Package Contents}

The package contains the following pieces of code:

\begin{enumerate}
 \item {\it formf} - a code to calculate some monopole and dipole form-factors,
 \item A modified version of {\it CHUCK3}, a code for DWBA calculations,
 \item {\it AngCor} - a code to calculate angular distributions
 \item {\it AverageAngCorResults} - a code to calculate the average (observed) angular distribution given the {\it CHUCK3} and {\it AngCor} outputs
 \item {\it TestCalculation.sh} - this is an example of how each part of the calculation should run. If you have compiled the codes correctly then you should be able to run this code and produce a test output.
\end{enumerate}

The components of the package are introduced separately in the following sections. In addition to the above components there are a number of shell scripts or other codes within the repository which may be used or modified as required in order to run various parts of the codes. These will be introduced in the relevant section. 

There is one example bash script to introduce now. This code is called {\it TestCalculation.sh} and it contains the commands to run each part of the whole package step-by-step. It is included for two reasons - first, so you can test to see if the code is working as expected and, second, so you may use it as a guide for running the calculations.

To download the code, one must run the command:\\
\mint[breaklines]{shell-session}|git clone --recurse-submodules https://github.com/padsley/AngCorPackage.git|

\noindent which should create a sub-directory called {\it  AngCorPackage} inside the directory where you run that command. Inside this new directory are some further subdirectories. At this point, please check that inside the {\it AngCorPackage} directory that a directory called {\it AngCorAveraging} exists and that it contains files. If it does not, please run the command:
\mint[breaklines]{shell-session}|git pull|
\noindent within the {\it AngCorAveraging} directory.

\section{Optical-model potentials}

Optical-model potentials are not part of this package. As a hint - if you are looking at $\alpha$-particle inelastic scattering then consider using the global optical-model potentials of Nolte, Machner and Bojowald \cite{PhysRevC.36.1312}. Other optical-model potentials are available in the literature; the \href{https://www-nds.iaea.org/RIPL-3/}{RIPL-3 database} is another useful source. At RCNP for $(\alpha,\alpha^\prime)$ reactions a double-folding potential is often used but that code is not included in the present package.

We would like to add other sources of optical-model potentials. If you are the owner of a code to calculate optical-model potentials and would be willing to allow your code to be included within this package or linked to as a submodule then please let us know.

\section{Form Factors}

This code to calculate form factors, {\it formf}, was written by M. Harakeh from KVI. Details about authorship and the origins of the various form factors may be found in {\it formf.for}. The form factors calculated are those of Harakeh and Dieperink \cite{PhysRevC.23.2329}, Orlandini et al. \cite{ORLANDINI198221} and Satchler \cite{SATCHLER1987215}.

Form factors for monopole and dipole states can be calculated using the code {\it formf}. This is necessary as {\it CHUCK3} only includes a first-order approximation of the isoscalar form factor for collective excitation that, for monopole and dipole transitions, are related to the centre-of-mass motion and thus cannot be related to an actual excitation of the nucleus. For this code, the input required is the $\alpha+$nucleus optical model potential. The outputs are form factors which may be used as inputs for {\it CHUCK3} calculations.

The first step here is to compile the code. To do this, go to the {\it formf} directory. Then go into the {\it cio} directory and run the command \mintinline{shell-session}|make|. Then go back up one directory (back to the {\it formf} directory) and run \mintinline{shell-session}|make| again.

An executable {\it formf} should now have been created. The code will then run line-by-line asking for the optical-model potential to be input piece-by-piece. This is worth doing to get to know how {\it formf} works. However, one can also run {\it formf} by giving an input file: \mintinline{shell-session}|./formf < input|.

An example input is:
\newline
\noindent pr\\
140\\
1\\
-18.7776\\
1.57\\
0.58815\\
30\\
.1\\
ex\\

An approximate translation of that input is:
\newline

\noindent pr - Proceed with the calculation\\
140 - Mass of the nucleus\\
1 - Type of the Woods-Saxon potential 1 = volume, 2 = surface\\
-18.7776 - Depth of the WS potential\\
1.57 - Reduced radius for the potential\\
0.58815 - Diffusiveness of the potential\\
30 - Maximum integration radius\\
.1 - Integration step size\\
ex - exit the code\\

Note that the maximum integration radius and the integration step size must both be identical between {\it formf} and {\it CHUCK3}. The user must verify that these values are identical.

It is generally easier to prepare these files beforehand: it is more convenient to use the code in this manner and it is useful to have the old input files available if you are trying to remember what you did some months previously.

\section{CHUCK3}

{\it CHUCK3} is a coupled-channels direct reaction written by Peter Kunz and modified by JR Comfort\footnote{The website of Peter Kunz appears to no longer be online. If you would like to access the {\it CHUCK3} guide, it is available \href{https://github.com/padsley/CHUCK3}{on Github}}. {\it CHUCK3} is a coupled-channel code to perform DWBA calculations. Its purpose in the current calculations is twofold. First, it is used to calculate the differential cross section as a function of scattering angle and second, it is used to obtain the substate distribution which is required for {\it AngCor}. To do this, a {\it CHUCK3} input file must be prepared. Be careful - these input files are read in with FORTRAN and are thus sensitive to whitespaces and which column quantities are aligned to. It is therefore often much easier to take an existing file and modify it rather than writing one from scratch.

The first step for using {\it CHUCK3} is to compile the code. This should be done by entering the {\it chuck} directory and running the command \mintinline{shell-session}|make|. This should create a {\it chuck} executable.

{\it CHUCK3} is detailed in the instruction manual in the {\it chuck} directory. Note that the version of {\it CHUCK3} in the present case is a modified version of the {\it CHUCK3} code and differs from Kunz's original {\it CHUCK3}.

An annotated input is given below - however, please do not try to use it to make an input yourself because \LaTeX\ has likely modified the formatting enough that this will not work. Instead, use the example inputs in {\it chuck/input} as templates. However, a brief note about the form factors - the code {\it formf} gives monopole and dipole form factors. For higher-order cases, one can describe the form-factor in a variety of fashions. However, for the purposes of {\it AngCor} it is possible to just use a form factor of one of the reaction potentials.

An example {\it CHUCK3} input for the excitation of the isoscalar giant dipole resonance in $^{48}$Ca by the inelastic scattering of 136-MeV $\alpha$-particles is given below and followed by a line-by-line description:\\

\noindent 11     23000     1    Ca48  136 MeV      ISGDR Excitation\\
100.    0.0    0.15\\
150  2  0 -2\\
0.1     30.\\
136.    4.      2.      48.     20.     1.4\\
  1  1\\
-1.     -100.7  1.25    0.78            -21.4   1.57    0.62 \\
-7.600  4.      2.      48.     20.     1.4\\
  2  2\\
-1.     -100.7  1.25    0.78            -21.4   1.57    0.62 \\
 -2  1  1  0  2  3  0  0 0.10\\
6.      222.49771.25    0.78            37.569061.570   0.62            1.00\\
7.      -967.3841.25    0.78            -247.8991.570   0.62            0.00\\
7.      84.911021.25    0.78            18.198681.570   0.62            2.00\\
-8.     4.71157 1.25    0.78            1.28178 1.570   0.62            1.00\\
            
\noindent 110000023000     1    Ca48  136 MeV      ISGDR Excitation\\
123456789012 - these numbers are here to help you line up the ICON values
Options and title for the file. The options are called ICON in the list. In this case, the options are:

\noindent ICON(1)=1 [read additional cards for this run],\\
\noindent ICON(2)=1 [according to {\it CHUCK3} documentation this is not used but the modified {\it CHUCK3} code says that this suppresses the printing of form factors if desired],\\
\noindent ICON(3-7) are blank or 0\footnote{In order to show the ICON values more clearly, 0s have been added for each blank.},\\
\noindent ICON(8)=2 [Print out scattering amplitudes],\\
\noindent ICON(9)=3 [Print out 3-cycle semilog plot of the differential cross section],\\
\noindent ICON(10-12)=0 [Don't use relativistic kinematics and turn off a couple of printing options]. \\ \\

\noindent 100.    0.0    0.15\\
Number of angles to calculate (100), starting at $\theta_\mathrm{cm} = 0.0$ in steps of 0.15 degrees.\\

\noindent 150  2  0 -2\\
Use 150 partial waves, with 2 channels the first having $J^\pi = 0^+$ and the second having $J^\pi = 1^-$. Note that the value for the spin is $2J$ and the sign gives the parity of the state. Therefore, a $J^\pi = 1^-$ state is written as $-2$, a $J\pi = 2^+$ state will be $+4$ etc.\\

\noindent 0.1     30.\\
Integration step size (0.1 fm) and maximum radius for the integration (you shouldn't need to change these but if you're worried about it, try using smaller step sizes and larger radii and seeing if anything changes. However, these values must be the same as those values used in the {\it formf} calculation.\\

\noindent 136.    4.      2.      48.     20.     1.4\\
Lab energy (136 MeV), the mass and charge of the projectile (4 and 2 for $^4$He), target mass and charge (48 and 20 for $^{48}$Ca) and reduced Coulomb charge radius (1.4 fm).\\

\noindent 1  1\\
This defines the channel (number 1), $J^\pi = 0^+$ in this example.\\
  
\noindent -1.     -100.7  1.25    0.78            -21.4   1.57    0.62\\
This defines an optical-model potential for the ingoing channel. The (-1) should be split into two parts: 1=a volume Woods-Saxon potential is being defined and the negative sign means that this is the final potential which will be read for this channel. -100.7 MeV, 1.25 fm and 0.78 fm are the depth, reduced radius and diffusivity of the real part of the Woods-Saxon potential, and -21.4 MeV, 1.57 fm and 0.62 fm are the equivalent parameters for the imaginary part.\\

\noindent -7.600  4.      2.      48.     20.     1.4\\
The definition of the second channel, $J^\pi = 1^-$ in this example. The excited state is at $E_x = 7.6$ MeV, the $\alpha$-particle and $^{48}$Ca definitions remain the same.\\

\noindent 2  2\\
Defines the second channel.\\
  
\noindent -1.     -100.7  1.25    0.78            -21.4   1.57    0.62\\
Again defines a volume Woods-Saxon potential, in this case identical to that given for channel number 1.\\

\noindent -2  1  1  0  2  3  0  0 0.10\\
This line (card) defines the coupling between the two channels. The coupling is from channel 1 to channel 2 (the final channel is given first). The coupling is only in one direction because it is a negative value (-2), this reduces the problem to a single-step DWBA calculation. If the final channel is a positve number then coupling in both directions is allowed. The next value (1) defines the initial channel. The next value (1) defines the orbital angular momentum of the transfer. The next value (0) is twice the value of the spin transferred. The next value is twice the total angular-momentum transfer (2). The next value (3) defines the form factor. The next value (0) is the order of deformation for the collective form factor if the previous value was 0. The next value (0) is a flag telling {\it CHUCK3} not to calculate Coulomb excitation; non-zero values will cause Coulomb excitation to be calculated. The final value (0.10) gives the $\beta$ for the calculation. The $\beta$ is the coupling strength, for details as to the definition of $\beta$, refer to the {\it CHUCK3} manual. \\
 
\noindent 6.      222.49771.25    0.78            37.569061.570   0.62            1.00\\
First part of the form factor definition. This is a volume form factor of the form $[V(r) + iW(r)] * r^n$. In this case, the parameters for the real and imaginary parts are 222.49771 MeV, 1.25 fm and 0.78 fm, and 37.56906 MeV, 1.570 fm and 0.62 fm respectively. The 1.00 at the end of the line is $n$, the exponent of the factor with $r$.\\

\noindent 7.      -967.3841.25    0.78            -247.8991.570   0.62            0.00\\
Second part of the form factor definition. This is a surface form factor with an $r^n$ component. For this component of the form factor $n=0$.\\

\noindent 7.      84.911021.25    0.78            18.198681.570   0.62            2.00\\
Third part of the form factor definition. This is a surface form factor with an $r^2$ component.\\

\noindent -8.     4.71157 1.25    0.78            1.28178 1.570   0.62            1.00\\
Fourth and final (the 8 which defines the type of the form factor is negative) part of the form factor definition. This is a second-order form factor with an additional $r$ dependence.\\

To run {\it CHUCK3}, one should do \mintinline{shell-session}|./chuck < input| to print the output to the screen or \mintinline{shell-session}|./chuck < input > output| to print the output to a file (called output). The output file is a long text file which prints out part of the status for the calculation. Another important file which is generated is the \mintinline{shell-session}|fort.2| file, containing the population of the substates for the scattering at a particular angle.

Please note that the cross section values are required for the averaging code and so you must use {\it CHUCK3} in the form \mintinline{shell-session}|./chuck < input > output| when running {\it CHUCK3} in preparation for doing averaged {\it AngCor} calculations.

One particular odd behaviour with {\it CHUCK3} that has been observed is that some calculations will not run unless you have a two blank lines (i.e. carriage returns) at the end of the input file. The reason for this is not entirely understood but if you can see that the outputs are all being read correctly before an error that looks like:\\
\begin{minted}[breaklines]{shell-session}
At line 287 of file chuck.for (unit = 5, file = 'stdin')\\
Fortran runtime error: End of file 
\end{minted}
consider adding a couple of additional lines to the end of the code and trying again. Alternativly, one can try using ICON(1)=9 for the next line which, in theory, should terminate the calculation at the end of the one being performed.

\section{AngCor}

{\it AngCor} was written by M Harakeh and LW Put, with modifications made by M Yosoi and RGT\footnote{These initials are given in the FORTRAN input file of {\it AngCor} but the full name is not given}.

\begin{figure}[phtb]
\centering
\begin{tikzpicture}

\draw[->] (0,0) -- (-3,0);
\draw[->] (0,0) -- (0,3);
\node at (-3.3,0) {$x$};
\node at (0,3.3) {$y$};

\draw[red,dashed,->] (0,0) -- (-2,2);
\draw[red] (-1,0) arc (180:135:1);
\node[red] at (-1.2,0.5) {$\phi_\alpha$};

\draw[blue,dotted,->] (0,0) -- (2,3);
\draw[blue] (-0.75,0.75) arc (135:56.31:1);
\node[blue] at (-0.2,1.28) {$\phi_d$};

\end{tikzpicture}
\vspace{2cm}
\begin{tikzpicture}
\draw[<-] (-2.598,1.5) -- (2.598,-1.5);
\draw[->] (0,-3) -- (0,3);
\draw[->] (-2.598,-1.5) -- (2.598,1.5);

\node at (-2.858,1.65) {$x$};
\node at (0,3.3) {$y$};
\node at (2.858,1.65) {$z$};

\draw[red,dashed,->] (0,0) -- (-0.8377311788,3.503113825);
\draw[red] (-0.2792437263,1.167704608) arc (103.45:19:1);
\node[red] at (0.7,1.3) {$\theta_\alpha$};
\draw[red] (-0.223394981,0.9341636866) arc (103.45:161:0.8);
\node[red] at (-0.8,1) {$\phi_\alpha$};

\draw[blue,dotted,->] (0,0) -- (1.441150837,1.664104021);
\draw[blue] (1.20095903,1.386753351) arc (49.10671396:103.4490799:1.834499236);
\draw[blue] (0.9607672244,1.109402681) arc (49.10671396:30:1.467599389);

\node[blue] at (1.25,1.1) {$\theta_d$};
\node[blue] at (0.5,2.1) {$\phi_d$};

\end{tikzpicture}
\caption{Schematic representation of the angles. (Left) Looking along the beam ($z$) axis defining the reaction plane relative to the laboratory coordinates, and defining the decay azimuthal angle, $\phi_d$, relative to the reaction plane. (Right) Out-of-plane view showing all of the axis definitions. Note that the decay azimuthal angle, $\phi_d$, is defined relative to the reaction lane of the scattered particle but the decay polar angle, $\theta_d$, is defined relative to the beam ($z$) axis. The axes are solid black lines, the scattered particle is a dashed red line and the decay particle is a dotted blue line.}
\label{fig:SchematicAngles}
\end{figure}
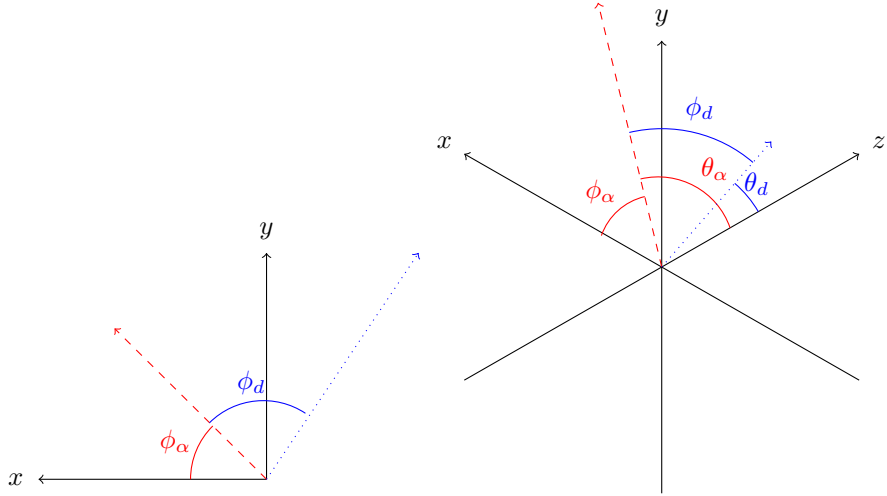

{\it AngCor} calculates the angular correlation function as a function of the polar decay angle ($\theta_{\mathrm{decay}}$) for a reaction involving the scattering of the ejectile at a particular polar scattering angle ($\theta_{\alpha}$ - note that this does not have to be an $\alpha$ particle - it is just a quirk of the naming convention as this code was originally written for use with $\alpha$-particle inelastic scattering reactions with the K600) and azimuthal decay angle ($\phi_{\mathrm{decay}}$). A schematic diagram showing the angles is provided in Figure \ref{fig:SchematicAngles}.

The first step in using {\it AngCor} is to compile the code. This should be done by running \mintinline{shell-session}|make| in the {\it angcor} directory. This should create some object (.o) files and an {\it angcor} executable.

Once this has been done, the {\it AngCor} calculations may be run. An example input file is given below, followed by an annotated discussion of the input. Note that because {\it AngCor} must be run for a large number of cases corresponding to various scattering and $\phi_\mathrm{decay}$ angles, some {\it c} files are included in the folder for generating the inputs as well as shell scripts for running all of the generated inputs and an awk script for combining the outputs into one large file.

An example {\it AngCor} input for the $^{24}$Mg($\alpha,\alpha^\prime\alpha$)$^{24}$Mg inelastic-scattering excitation followed by $\alpha$-particle emission is shown below. The same warning as above about the {\it CHUCK3} files given above also applies here - copying and pasting this input from \LaTeX into a file will likely fail due to the formatting having gone wrong. In addition, there needs to be a new {\it AngCor} input file created for each scattering and decay angle pair so manually creating these inputs is prohibitively time consuming. An example of how to generate these inputs automatically is given after a description of the {\it AngCor} inputs. \\

\noindent 1- - state of 24Mg, Ex=9.305 MeV\\
\noindent 3,0,1,0,1,0,1,0,0,0,0,\\
\noindent 150,0,0,0,2,0,1,\\
\noindent 1,1,2,\\
\noindent 0,0,0,\\
\noindent 3.6,134.,0.,0.,0.,\\
\noindent 0.,0.,0.,0.,\\
\noindent 1,181,\\

\noindent A line-by-line interpretation of the {\it AngCor} input is:\\

\noindent 1- - state of 24Mg, Ex=9.305 MeV\\
\noindent Line containing title information\\

\noindent 3,0,1,0,1,0,1,0,0,0,0,\\
\noindent 1,2,3,4,5,6,7,8,9,n,m, (This line added to guide the eye for the ICON values.)\\
Another set of ICON inputs. In this case, the ICON values are:\\
ICON(1) = 3 - particle-fission fragment angular correlation.\\\
ICON(2) = 0 - do not print information from $RHOKQ$ and $RHOLSJ$ subroutines.\\
ICON(3) = Calculate $\bar{RHO}_{KQ}$ from $RHO_{KQ}(LSJ,L^\prime S^\prime J^\prime)$.\\
ICON(4) = 0 - Only one intermediate state is excited.\\
ICON(5) = 1 - Transition amplitudes are read from a {\it CHUCK3} file. This file is called {\it fort.2} and should be in the directory where the {\it AngCor} code is run.\\
ICON(6) = 0 - use new CHUCK information.\\
ICON(7) = 1 - Print the $D$ values from {\it CHUCK3}. The $D$ values are the transition amplitudes from {\it CHUCK3}, for details as to the definition of the $D$ values, consult the {\it CHUCK3} manual.\\
ICON(8) = 0 - Print angular correlation values.\\
ICON(9) = 0 - the z-axis is the beam axis for cases where the m-state populations are obtained from CHUCK (as in the present case).\\
n is NC in the {\it AngCor} documentation, and defines the number of cascades in the case of $\gamma$-ray decay.\\
m is M1 in the {\it AngCor} documentation. It defined the number of experimental data for the minimisation of the angular correlation. We tend not to use this option.\\

\noindent 150,0,0,0,2,0,1,\\
This card defines the reaction spins. The number of partial waves (150), twice the spin of the projectile (0), twice the angular momentum of the target (0), twice the angular momentum of the ejectile (0), twice the angular momentum of the Nth (1st) intermediate state populated in the reaction (2), twice the spin transfer for that state (0) and a scaling factor defining the ratio of excitation to the various states which is defunct in the present case as {\it CHUCK} is used for the calculation.\\

\noindent 1,1,2,\\
This card defines the transfer of the angular momenta. The first value is the number of separate angular-momentum transfers (1), the second value is the value of the 1st angular-momentum transfer (1) and the final value is twice the value of the 1st angular momentum transfer (2). For cases where there are multiple stages to the excitation then the first value will be equal to the number of transfers (NLTR) and then there must be NLTR values given for the orbital and total angular-momentum transfers.\\

\noindent 0,0,0,\\
Decay channel parameters. These can differ for particle decays, $\gamma$-ray decays or fission decays. In this case we are using a fission file as an input and so these values are twice the projection of the initial spin on the symmetry axis, and twice the projection of the final spin on the symmetry axis. The final 0 does not do anything and can be ignored.\\

For a $\gamma$-ray decay, this line may look something like:\\
1,0,0,\\
In this case, there should be a line like this for every $\gamma$-ray transition. The first value (1) is the lowest multipolarity in the transition, this will be mixed with multipolarity of one spin higher. The next value (0) is twice the spin value of the state to which the $\gamma$-ray transition proceeds. The final value (0) is the minimum angle of arctan($\delta$), where $\delta$ is the mixing ratio between the two multipolarities.

At this point, if looking at $\gamma$ ray decays the attenuation coefficients would be defined for $L=2$, $4$ and $6$ multipole terms in the angular correlation. The values we use for this are:
0.98885,0.963146,0.92366,\\
Updated information on the origin of these terms will be added to the guide when available,

\noindent 3.6,134.,0.,0.,0.,\\
This card defines the angles for the interaction. The first value (3.6) defines the centre-of-mass polar angle of the scattered particle with respect to the beam axis. The second value (134) is the azimuthal angle of the decay particle relative to the reaction plane. A value of 0 represents decay in the reaction plane on the same side of the beam axis as the ejectile/scattered particle. The next value is the recoil angle, but this is only used if ICON(9)=1. The final two values may be ignored.

The angles defined above (3.6 and 134 degrees) will change for each {\it AngCor} input file corresponding to different scattering and decay angles. The purpose of the generation code discussed below is to make all of the {\it AngCor} files for these different angle pairs automatically. \\

\noindent 0.,0.,0.,0.,\\
This card defines the ejectile-detector opening angle but we do not use this part of the code in the particle decay section.

\noindent 1,181,\\
These define the angle step and the maximum angle for the angular correlation function.\\

One peculiarity of {\it AngCor} that has been noted (for example, for the work described in Ref. \cite{PhysRevC.95.031302}) is that the particle-particle coincidence option does not seem to produce any output. We have obtained good results using the particle-fission option as an alternative to the particle-particle coincidence option.

An example of the generating code for {\it AngCor} inputs is {\it make\_input\_010.c}. This code makes {\it AngCor} inputs for reactions from a $J^\pi = 0^+$ initial state, to a $J^\pi = 1^-$ excited state followed by an $E1$ decay to a $J^\pi = 0^+$ final state. To run the code, one must first compile it using the command:\\ \mint{shell-session}|g++ make_input_010.c -o make_input_010|.
\\The code is then run with the command:\\
\mint{shell-session}|./make_input_010.|
\noindent This should produce a lot of output files with {\it .com} extensions. The filenames include the $\theta_\alpha$ and $\phi_\mathrm{decay}$ for that particular {\it AngCor} input file.

Once each of the {\it AngCor} inputs has been generated in the {\it input} directory, the {\it AngCor} calculations may be performed. This may be done using shell scripts like those found within the {\it input} directory. An example can be found in {\it DoAngCorPR244J\_1.sh} and easily modified to take into account changes to file names. This step may produce a number of errors about IEEE standards - as far as we can tell these are warnings about numerical precision in the compiler but for most of the examples that we have run, the final results are acceptable.

Running all of the {\it AngCor} files should take the input {\it .com} files in the {\it input} directory and produce a large number of {\it .lis} files in the {\it output} directory. 

The {\it make\_final.sh} shell script takes the {\it .lis} outputs and combines them into one file. This is done using the {\it sort.awk} script (more on this in a moment) invoked within the {\it make\_final.sh} script. The final output file is comprised of four columns: $\theta_\alpha$, $\theta_\mathrm{decay}$, $\phi_\mathrm{decay}$ and $W(\theta)$. This is the combined {\it AngCor} output for all of the various possible polar scattering angles and decay angles. This output file will be used in the averaging portion of the code.

Note that the {\it sort.awk} file which does the combination of the various runs is an inflexible tool. It uses substrings to search for the relevant angles ($\theta_\alpha$ and $\phi_\mathrm{decay}$) for each of the different inputs. Therefore, if the number of characters in the names of the output files is change then the size and starting character for the substrings in {\it sort.awk} will also need to change. Instructions for how to code in awk are beyond the scope of this document but may be easily found online. Additionally, one can add print statements to the awk file to check to see how the substrings are being interpreted. The version of the {\it sort.awk} file which is currently in the repository is designed to use files with names of the form {\it input\_PR244\_J\_1\_1.2\_25.com}.

If you are willing to make an updated version of the script to make the final {\it AngCor} output file please feel free to do so and we can include it in the package.

\section{Averaging the AngCor result}

This part of the code, found in the folder {\it AngCorAveraging} was written by P. Adsley based on an archetype code of Vera Derya.

A working ROOT distribution is a prerequisite for using this code. Instructions for how to do this may be found on \href{https://root.cern.ch/building-root}{the ROOT website}.

In many cases, the scattering reactions relevant for the current work use finite-acceptance magnetic spectrometers. This means that there is a range of angles which can be populated in the reaction. Especially in the $0-2$ degree range for the K600 aperture, this can result in a range of recoil angles over tens of degrees for the recoil. Therefore, the results of the {\it AngCor} calculations need to be smeared over the finite aperture size of the spectrometer.

In order to run this code, one needs to first modify the code and then to compile it. The modifications are mainly in {\it AverageAngCorResults.h}.

\begin{enumerate}
\item Ensure that the correct number of points are set for $\theta_\alpha$ (\emph{NumberThetaAlphaPoints}). The easiest option is to make it the same as the number of angles used in the {\it CHUCK3} calculation.

\item The number of {\it CHUCK3} angles (\emph{NumberOfCHUCK3Angles}) should be set to be the number of angle points which one used when running the {\it CHUCK3} code. If this is not done then it tends to result in the code giving odd results because the angles do not match.

\item Another variable -- \emph{ThetaAlphaStartAngle} -- is provided for setting the initial angle for the DWBA calculations if for some reason the calculations were performed starting at an angle greater than 0\textdegree.

\item The value of \emph{DeltaThetaAlpha} must be set to be the size of the increment on the angles in the {\it CHUCK3} input file.
\item The number of $\phi_\alpha$ points (\emph{NumberPhiAlphaPoints}) can remain 360.
\item The number of $\theta_\mathrm{decay}$ points (\emph{NumberThetaDecayPoints}) should be the same as the number of points used in the {\it AngCor} input file. This is probably 180 unless you changed something.

\item The number of $\phi_\mathrm{decay}$ points (\emph{NumberPhiDecayPoints}) must also be set. This is 180 and is unlikely to change from 180.
\end{enumerate}

If some of these values are set incorrectly then it is very possible that either the angular correlation produced will be incorrect or alternatively that the code will crash because the code will go beyond the end of an array while trying to store the cross section or {\it AngCor} data.

To compile, one needs to run the command:

\begin{minted}[breaklines]{shell-session}
g++ AverageAngCorResults.cpp -o AverageAngCorResults `root-config --cflags --libs` 
\end{minted}

The code requires three arguments to run. These may be found by running \mintinline{shell-session}{./AverageAngCorResults} with no arguments which will return the output:\\
\mint[breaklines]{shell-session}{usage: mcerr <filename for cross section> <filename for AngCor results> <filename for output>}.
\noindent The file for the cross section will be the file which was created in the command:\\ \mint[breaklines]{shell-session}{./chuck3 < input > output.}
\noindent For the file providing the {\it AngCor} results use the combined file created when running the \mintinline{shell-session}{make_final.sh} shell script. Note that one needs to also give the path for the input files if they are not within the current directory.
 
To get the angular correlation function in the centre-of-mass frame of the colliding system one can use the \mintinline{shell-session}{TTree} which was output in the ROOT file. In order to plot the ACF, one should run a command of the form:
\mint[breaklines]{cpp}{AngCorData->Draw("ThetaDecayCM>>hW(181,0,180)","Weight*(CUTS)");}

\begin{figure}[phtb]
 \includegraphics[width=\textwidth]{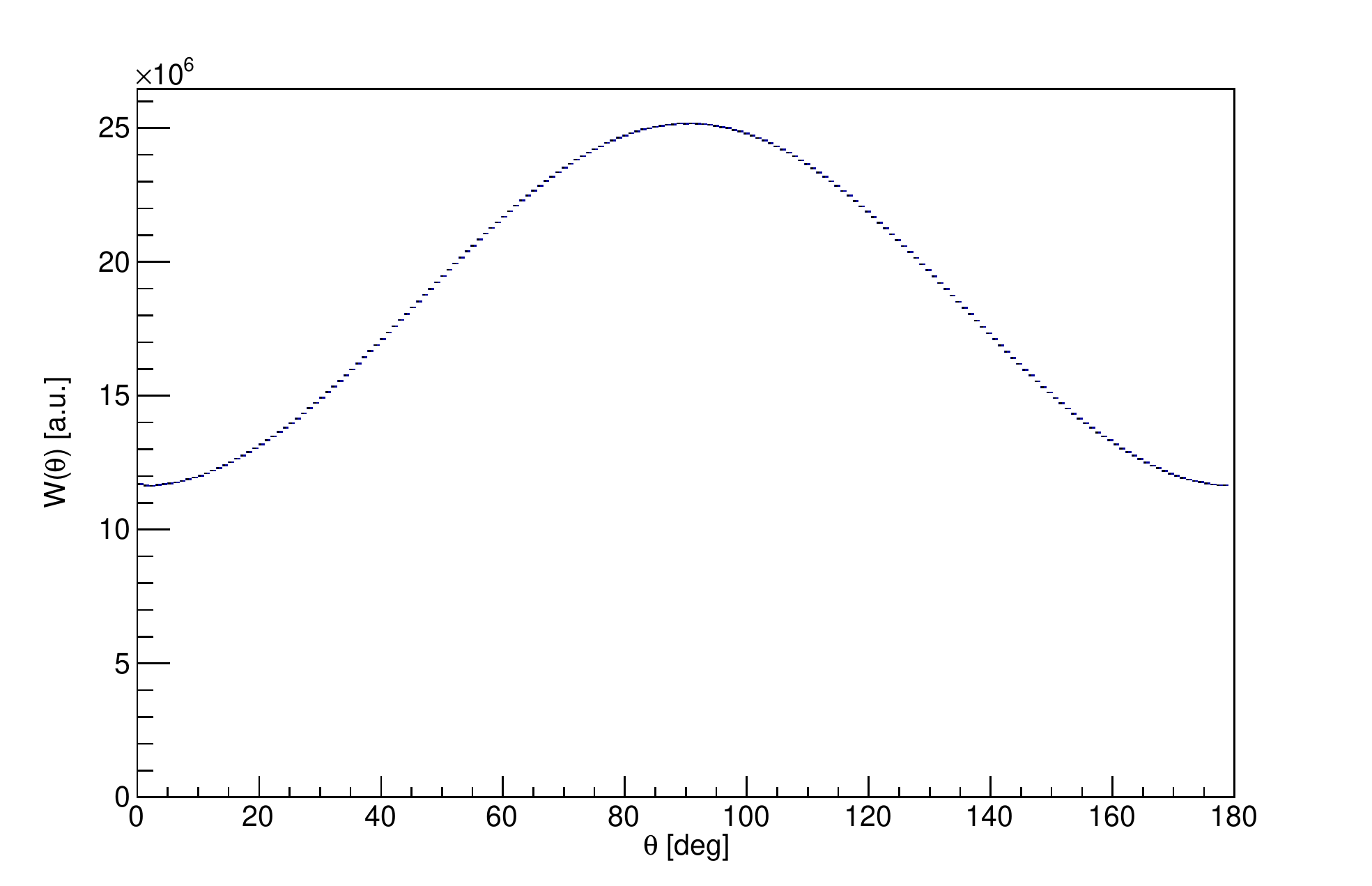}
 \includegraphics[width=\textwidth]{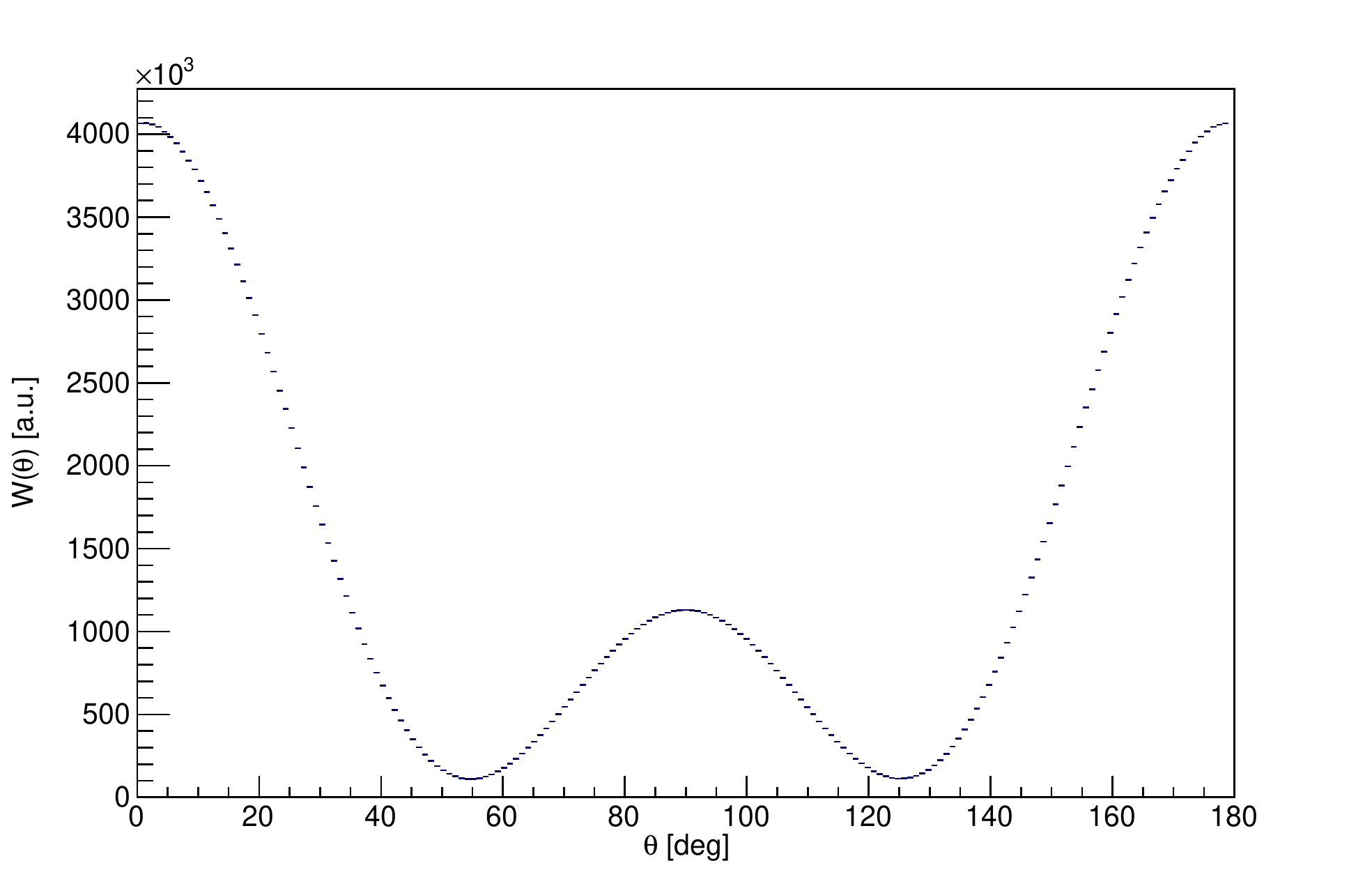}
  \caption{Angular correlation functions of $\alpha$-particle decays to the ground state of $^{20}$Ne for (top) $J^\pi = 1^-$ and (bottom) $J^\pi = 2^+$ states populated in the $^{24}$Mg($\alpha,\alpha^\prime$)$^{24}$Mg reaction.}
  \label{fig:AlphaEmission}
\end{figure}

\begin{figure}[phtb]
    \centering
    \includegraphics[width=\textwidth]{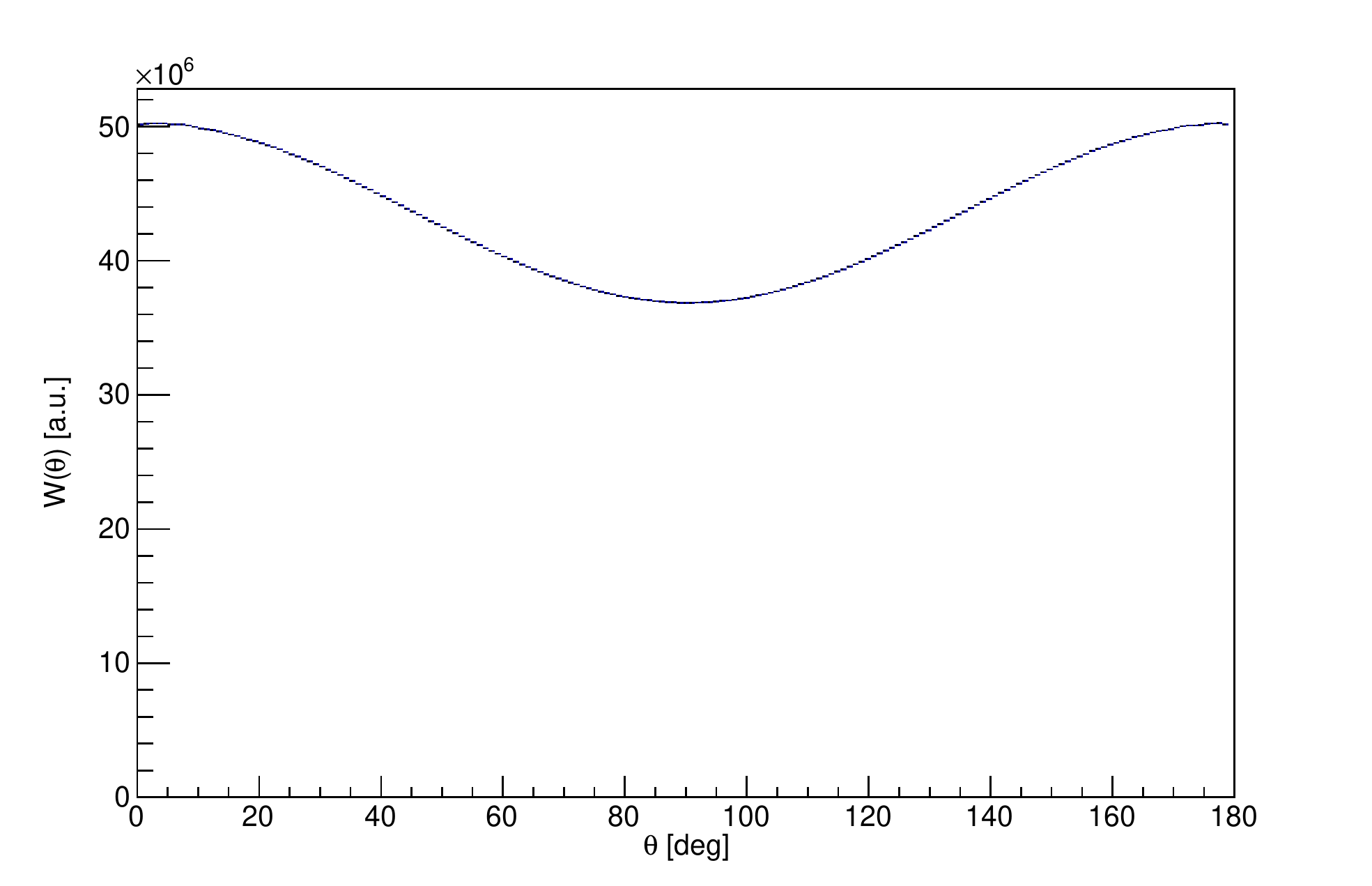}
    \includegraphics[width=\textwidth]{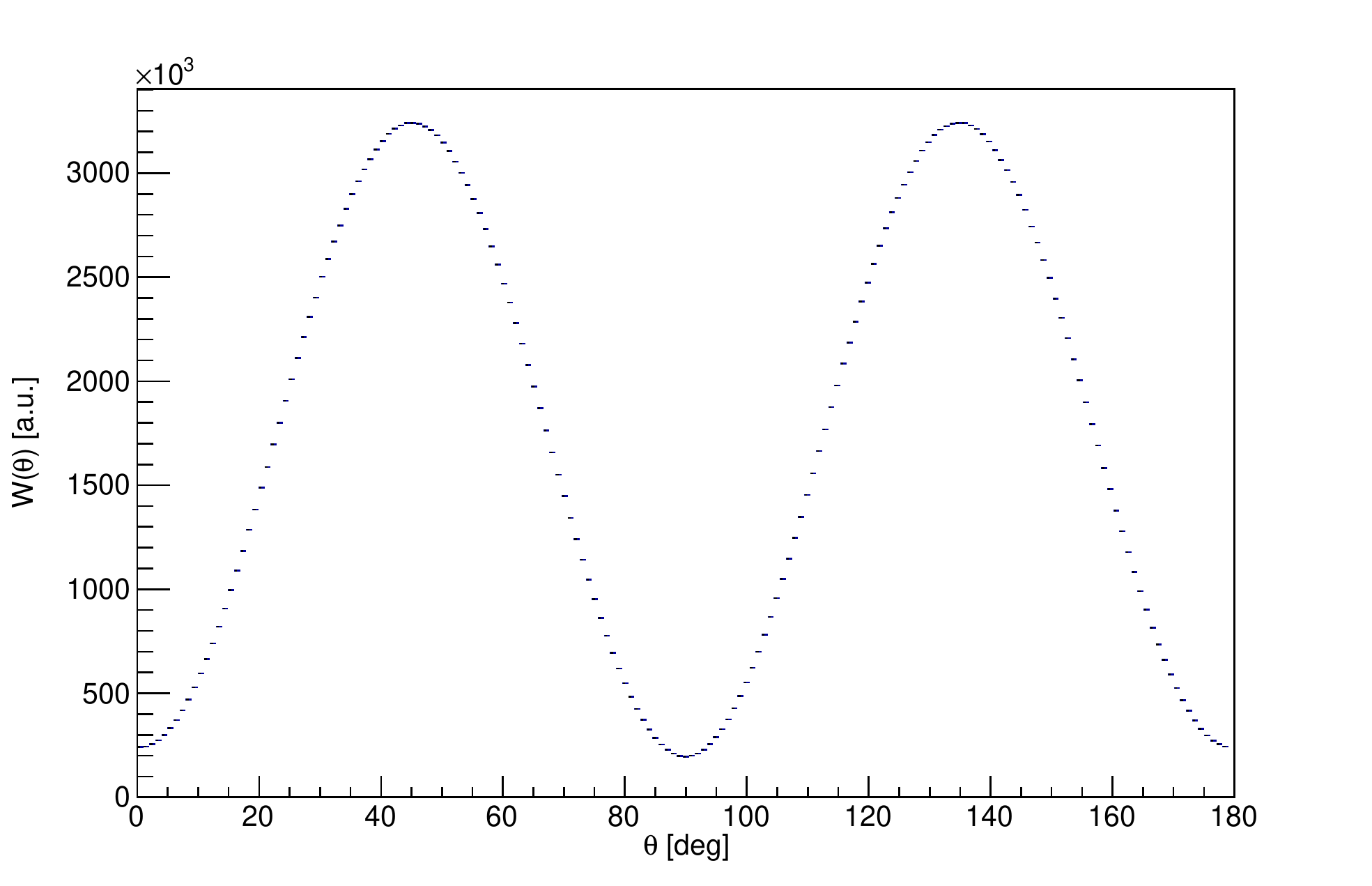}
    \caption{Angular correlation functions of $\gamma$-ray decays to the ground state of $^{24}$Mg for (top) $J^\pi = 1^-$ and (bottom) $J^\pi = 2^+$ states populated in the $^{24}$Mg($\alpha,\alpha^\prime$)$^{24}$Mg reaction.}
    \label{fig:GammaEmission}
\end{figure}

This will weight the $\theta_{\mathrm{decay}}$ distribution with the correct weight which is derived from the information on the differential cross section from {\it CHUCK3} and the angular correlation functions from {\it AngCor}. An example ({\it GenerateOutputAngularCorrelation.cpp}) of how to do this is given in the {\it AverageAngCorResults} directory. The resulting plot should look something like Figures \ref{fig:AlphaEmission} and \ref{fig:GammaEmission}.

Note that, at the moment the code does {\bf not} provide information on the lab angles of the decaying particles. The calculation of this value is a work in progress and will be added when available.

Note also that the averaged angular correlations are not normalised at the present time. However, this does not change the shape of the angular correlation function. Normalisation is fairly simple to accomplish in ROOT.

\section{Acknowledgements and Thanks}

None of this work would have been possible without Muhsin Harakeh, both for providing the codes in the first place and for allowing us to put them online. Any mistakes found in this brief guide are solely the responsibility of the authors.

\section{Change Log}

This section lists notable changes made to the codes in the {\it AngCorPackage} repository.

\begin{enumerate}
    \item October 2019: PA fixed the BINOM routine in {\it AngCor}. This code had a bug due to gfortran not initialising the array to be filled with 0s, which in turn caused the computation of the binomial coefficients to fail. For details, see \href{https://github.com/padsley/AngCorPackage/issues/3}{this issue on the Github repository}
    \item November 2019: PA added a new {\it TestCalculationGamma.sh} script which should do angular correlations for $E1$ and $E2$ decays following $^{24}$Mg($\alpha,\alpha^\prime$)$^{24}$Mg reactions.
\end{enumerate}

\bibliographystyle{unsrt}
\bibliography{AGuideToAngCor}
\end{document}